\documentclass[aps,prl,twocolumn,groupedaddress,showpacs,showkeys]{revtex4-1}
\usepackage{graphicx,color}
\usepackage{amsmath,amssymb,latexsym}
\usepackage[greek,english]{babel}
\usepackage{dcolumn}
\usepackage{bm}
\usepackage{float}
\usepackage[normalem]{ulem}

\usepackage{color}

\begin{document}

\title{Single-file escape of colloidal particles from
 microfluidic channels}

\author{Emanuele Locatelli}
\email{emanuele.locatelli@univie.ac.at}
\altaffiliation {present address: Faculty of Physics, University of Vienna, Boltzmanngasse 5, A-1090 Vienna, Austria}
\affiliation{
Dipartimento di Fisica e Astronomia `G. Galilei' - DFA and Sezione CNISM,
Universit\`a di Padova,
Via Marzolo 8, 35131 Padova (PD), Italy
}

\author{Fulvio Baldovin}
\email{baldovin@pd.infn.it}
\affiliation{
Dipartimento di Fisica e Astronomia `G. Galilei' - DFA, Sezione INFN and Sezione CNISM,
Universit\`a di Padova,
Via Marzolo 8, 35131 Padova (PD), Italy
}

\author{Enzo Orlandini}
\email{orlandini@pd.infn.it}
\affiliation{
Dipartimento di Fisica e Astronomia `G. Galilei' - DFA, Sezione INFN and Sezione CNISM,
Universit\`a di Padova,
Via Marzolo 8, 35131 Padova (PD), Italy
}

\author{Matteo Pierno}
\email{matteo.pierno@unipd.it}
\affiliation{
Dipartimento di Fisica e Astronomia `G. Galilei' - DFA and Sezione CNISM,
Universit\`a di Padova,
Via Marzolo 8, 35131 Padova (PD), Italy
}

\author{Yizhou Tan}
\affiliation{Cavendish Laboratory, CB30HE, Cambridge, UK
}

\author{Stefano Pagliara}
\email{s.pagliara@exeter.ac.uk}
\altaffiliation{present address: Department of Biosciences, College of Life and Environmental Sciences, University of Exeter, EX44QD, Exeter, UK}
\affiliation{Cavendish Laboratory, CB30HE, Cambridge, UK
}

\date{\today}

\begin{abstract}
{Single-file diffusion is a ubiquitous physical process exploited by
  living and synthetic systems to exchange molecules with their
  environment. It is paramount quantifying the escape time needed for
  single files of particles to exit from constraining synthetic
  channels and biological pores. This quantity depends on complex
  cooperative effects, whose predominance can only be established
  through a strict comparison between theory and experiments. By using
  colloidal particles, optical manipulation, microfluidics, digital
  microscopy and theoretical analysis we uncover the self-similar
  character of the escape process and provide closed-formula
  evaluations of the escape time. We find that the escape time scales
  inversely with the diffusion coefficient of the last particle to
  leave the channel. Importantly, we find that at the investigated
  microscale, bias forces as tiny as $10^{-15}\;{\rm N}$ determine the
  magnitude of the escape time by drastically reducing interparticle
  collisions. Our findings provide crucial guidelines to optimize the
  design of micro- and nano-devices for a variety of applications
  including drug delivery, particle filtering and transport in
  geometrical constrictions.}
\end{abstract}



\maketitle Living and synthetic systems exploit a variety of pores or
channels at the micro- and nano-scale, to transport particles and
molecules \cite{Bres:2013,Dekk:2007}. When the pore or channel cross
section is close to that of the particles, these are no longer able to
pass each other, a phenomenon known as single-file diffusion (SFD).
SFD plays a role in numerous processes such as the diffusion of ion or
water molecules in transmembrane proteins
\cite{Hodg:1955,Benz:1987,Jens:2002,Rasa:2008,Grav:2013,Port:2013},
the diffusion of adsorbate molecules in zeolites
\cite{Kukl:1996,Hahn:1996,Karg:2014}, water diffusion in nanotubes
\cite{Das:2010}, colloidal particles diffusion in one-dimensional (1D)
channels \cite{Lutz:2004,Wei:2000,Lin:2005}, drug delivery through
nanofluidic devices \cite{Yang:2010}, protein sliding along DNA
\cite{Li:2009}, charge carrier migration in polymer and superionic
conductors \cite{Khar:2010,Rich:1977}.  SFD is a fascinating process,
since it does not obey Fick's laws. As the sequence of particles
remains unaffected over time, anomalous behavior characterizes in fact
SFD systems.  Specifically, it has been shown that single-file
interactions imply that the mean square displacement (\textrm{MSD}) of
a tagged particle scales as ${t}^{1/2}$ \cite{Harr:1965}.\\ \indent
Recently, a wealth of theoretical approaches to analyze SFD have been
developed
\cite{Rode:1998,Chou:1999,Vase:2002,Koll:2003,BereC:2005,Marc:2006,Liza:2008,Talo:2008,Bark:2009,Zilm:2009,Cent:2010,Flom:2010,Khar:2010,Ryab:2011,Suar:2013,LocaA:2015}
and the dependence of the \textrm{MSD} on time has been experimentally
investigated for many SFD systems.  (\textit{i}) At the nanoscale with
measurements on zeolites \cite{Hahn:1996,Kukl:1996} and single-walled
carbon nanotubes \cite{Das:2010} via nuclear magnetic resonance. (\textit{ii})
At the micro scale, on colloidal suspensions confined in circular
trenches \cite{Wei:2000}, in 1D circular channels created by means of
scanning optical tweezers \cite{Lutz:2004}, in narrow straight grooves
\cite{Lin:2002, Lin:2005} and in narrow microfluidic channels
\cite{Siem:2012}. (\textit{iii}) At the millimeter scale, on macroscopic
charged metallic balls electrostatically interacting and confined in a
circular channel \cite{Coup:2006}. 
So far, both experimental and theoretical approaches have been 
focused on the temporal dependence of the \textrm{MSD} of the particles inside the channels.  
On the contrary, little is known on the escape process of a single-file of particles out of 
a narrow channel, a key issue in the analysis of
diffusive transport in compartmentalized systems~\cite{Dagd:2012}. \\ \indent 
In this Letter we address this problem with a combined experimental and theoretical investigation that allows us to dissect the single contributions of the different cooperative effects involved in the escape process. The experiments are based on colloidal particles in microfluidics~\cite{Xu:2010,Wond:2011,Lin:2014,Blei:2007}, holographic optical tweezers (HOTs)~\cite{Curt:2002,Grie:2003,Padg:2011N,Padg:2011L} and digital video microscopy \cite{Croc:1996}.
This setup allows us
to monitor the position of colloidal particles in single file within 
arrays of microfluidic channels with different lengths ~\cite{Pagl:2013,Dett:2014,DettP:2014,Pagl:2014}. We measure the survival
probability for the last particle to leave each channel and the mean
escape time needed for all the particles to leave the channel. We compare these measurements with analytical predictions developed on the basis of the Reflection Principle Method~\cite{Rode:1998,LocaA:2015}. This allows us to successfully validate closed-formulas for the estimation of the mean escape
time which can be used for quantitative assessment in widespread single-file
conditions, on both living and synthetic model systems \cite{Dekk:2007,Jens:2002}. 
We find that (\textit{i}) the escape process of $N$
particles can be entirely described in terms of the survival
probability of the last particle to leave the channel; (\textit{ii})
the escape time scales inversely with the diffusion coefficient of a
single particle in the channel; (\textit{iii}) bias forces as tiny as
$10^{-15}\;{\rm N}$ determine the magnitude of the escape time by
drastically reducing interparticle collisions and switching-off
excluded-volume effects.\\ 

\begin{figure}
\includegraphics[width=7cm]{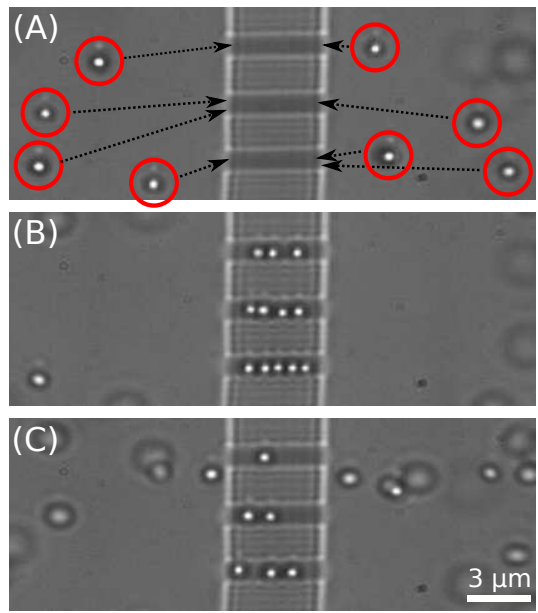}
\caption{(Color online) Filling and escape processes of colloidal
particles from an array of microfluidic channels.  (A) Two 3D baths
are filled with freely diffusing particles.  The baths are connected
by an array of microfluidic channels with similar
cross section and length $L_c$ of 4.7 $\mu$m. Eight optical traps are generated via holographic
optical tweezers and positioned in the 3D baths. Trapped particles
(highlighted with circles) are independently dragged and accurately
positioned in the microfluidic channels (dotted arrows depict the
dragging trajectories till the channel entrances). (B) Afterward, four
more optical traps are generated in the baths and used to drag four
more particles in the channels so that an array of \textit{N}=3, 4 and
5 single-file particles is generated in the top, central and bottom
channel, respectively. Particles are then released by switching off
the trapping laser at \textit{t}=0. (C) Exemplar snapshot of the
escape processes at \textit{t}=18 s. The escape processes are followed
till all the particles in the array of channels have escaped to the 3D
baths.}
\label{fig:fig1}
\end{figure}

\begin{figure}[!h]
\includegraphics[width=8.5 cm]{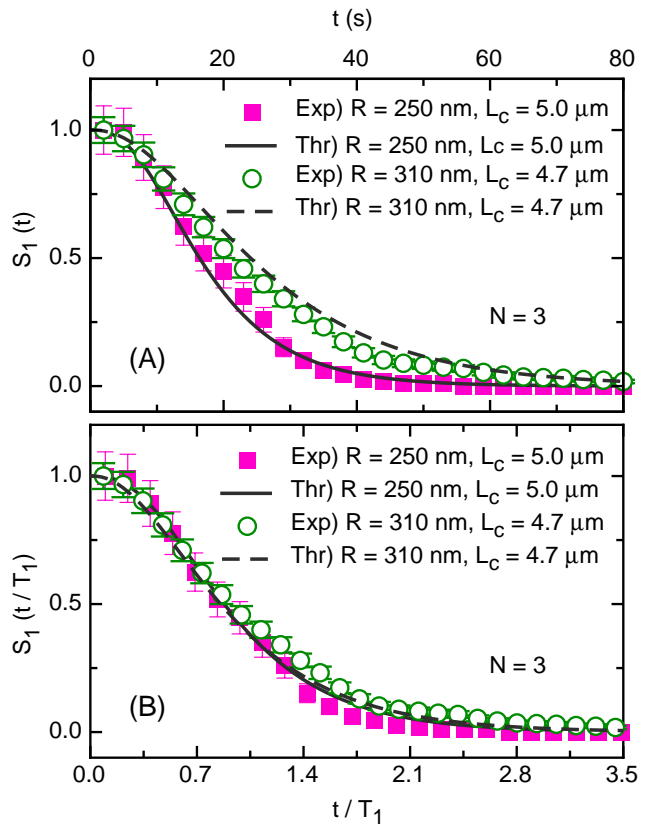}
\caption{(Color online).  (A) Time dependence of the escape-process
survival probability $S_1(t)$ of $N$=3 particles in
single-file. Squares and circles show the experimental results. 
Solid and dashed curves refer
to theoretical predictions calculated according to
Eq.~(\ref{eq:newfrompre}). (B) Dependence of $S_1$ on $t$/$T_1$ for
the same quantities plotted in panel (A).
Further plots are reported in the Supplementary Material~\cite{Loca:2015}.}
\label{fig:fig2}
\end{figure}

\indent Microfluidic devices consisting of two 3D baths with a depth of 16 $\mu$m separated
by a polydimethylsiloxane (PDMS) barrier and connected by an array of
microfluidic channels were fabricated as previously reported \cite{Pagl:2011,PaglE:2014}. The channels have cross-section dimensions
close to 1 $\mu$m, and length $L_c$ of 4.7 (Fig.~\ref{fig:fig1}A),
5, 6, 7, 9.6, or 12 $\mu$m. For full details about the channel
geometries, see Table S1 in the Supplementary Material~\cite{Loca:2015}. The baths are
filled with polystyrene spherical particles with radius
\textit{R}=(252.5$\pm$4) or (310$\pm$5) nm dispersed in a 5 mM KCl
salt solution. The Debye length associated with the particles is around 6 nm, much
smaller than the particle radii, thus justifying the assumption of
hard-sphere interactions in the theoretical model below. We use a
custom made HOTs setup \cite{Pagl:2013} to generate multiple
optical traps in the 3D baths, where particles freely diffuse
(Fig.~\ref{fig:fig1}A and Video S1). Upon trapping, particles are
independently dragged and positioned in the microfluidic channels with
an accuracy down to 100 nm (Fig.~\ref{fig:fig1}B), generating an array
of single-file particles with different initial numbers \textit{N}= 3,
4 and 5 in each channel (Fig.~\ref{fig:fig1}B). Once the array is completed, 
all particles are released at the same time by switching
off the trapping laser.
The escape processes from the
different channels are followed at all times. Fig.~\ref{fig:fig1}C reports an
exemplary snapshot of the escape processes after 18 seconds from the
release instant.  
Each
escape process is repeated at least 50 times. Overall, we investigate
escape processes from 13 different channels involving the trapping
and escape of more than four thousand particles (Table S1). We track
the position of each particle in each channel at all times by using a
custom-written Interactive Data Language routine based on a standard
particle tracking approach \cite{Croc:1996}. 
In each experiment we measure $J_{r-l}$ (see Table~S1 and Figs.~S1A, S1B),
the average difference in the number of particles exiting from each
channel to the right and left baths~\cite{Chodera:2011,PaglE:2014}. 
This allows us to identify the intensity of
any external bias force $F_e$ down to the scale of $10^{-15} \;{\rm N}$
and its contribution to the escape process via adimensional ratio 
$
(F_{e}\,L_c)/(k_{B}T)
$
(see Supplementary Material~\cite{Loca:2015}).\\
 \indent We measure the escape time $\tau$, i.e. the time
required for all the particles to exit from the channel to the
baths. Detailed information about the escape process can be obtained
in terms of the survival probability, experimentally defined as:\\
\begin{equation}
S_1(t) \equiv\frac{M_{t}}{M}= \mbox{Prob} \Big\{ \tau > t \Big\},
\end{equation}
where \textit{M} is the number of repeats of the same escape process
and $M_t$ is the subset
of repeats for which $\tau$ is larger than
\textit{t}. The corresponding errors are evaluated as
$\sqrt{M_t}/M$. 
From $S_1(t) $ one can estimate the mean escape time $T_1 =
\int_0^{+\infty}dt\,S_1(t)$. Note that both $S_1(t) $ and $T_1$
depend on several experimental parameters, namely the length of the
channel $L_c$, the number of initial particles $N$, and the width $L_0
\le L_c$ of the distribution of the initial positions of the particles
when these are released by switching off the trapping laser (Fig.~S2).
Moreover, a systematic external bias force $F_e$ can also
affect $S_1(t) $.  \\ 
\indent In order to gain information on the different physical mechanisms contributing to the escape process, we
carry out an analytical evaluation of $S_1(t) $ and $T_1$ based on the
Reflection Principle Method~\citep{Rode:1998,LocaA:2015} (full
details are reported in the Supplementary Material~\cite{Loca:2015}). 
The starting quantity is the conditional
probability $S_n(t|m,L_c,L_0,\Gamma)$.  This is the survival
probability of at least $n$ particles, given that at time $t=0$ $m\geq
n$ were uniformly distributed within the interval $[-L_0/2,L_0/2]$ of
a channel of total length $L_c\geq L_0$. The parameter $\Gamma\equiv
F_e/ 2 k_B T$ quantifies the bias. 
With $n=1$ and
$m=N$, $S_n(t|m,L_c,L_0,\Gamma)$ corresponds to the experimental
observable $S_1(t) $.  
In the limit of point-like particles the following relation
holds true \cite{LocaA:2015}:
\begin{equation}
1-S_1(t|N,L_c,L_0,\Gamma)
=
[1-S_1(t|1,L_c,L_0,\Gamma)]^N\;.
\label{eq:newfrompre}
\end{equation}
The analytical expression for the single-particle survival probability
in the presence of a bias, $S_1(t|1,L_c,L_0,\Gamma)$, is reported in
Eq. (6) of the Supplementary Material~\cite{Loca:2015}. By inserting this formula in
Eq.~\eqref{eq:newfrompre}, we obtain an analytical expression for
$S_1(t)$.  Importantly, Eq.~\eqref{eq:newfrompre} implies that the
multi-particle escape process can be mapped onto the escape process of
many independent particles.  Indeed the multi-particle escape from the
channel can be considered as a collective process where the order in
which particles escape from the channel can be neglected. Therefore,
in our model we do not tag any of the particles, thus neglecting
collisions (equal-mass particles exchange their velocities in 1D
elastic collisions) while assuming independent point-like particles.
Intuitively, while collisions hamper the diffusion of the particles in
the center of the single file, they simultaneously contribute to 
 expel those at the edges of the channel, 
these two effects canceling each other. As a
consequence the $S_1(t)$ values calculated according to
Eq.~\eqref{eq:newfrompre} (lines in Fig.~\ref{fig:fig2}A) favorably
compare with the experimentally measured values (symbols in
Fig.~\ref{fig:fig2}A).  Remarkably, by rescaling $t$ to $t/T_1(N,L_c)$
both the experimental data and the analytical expressions collapse
onto the same curve (Fig.~\ref{fig:fig2}B and Fig. S4). This suggests
that $T_1$ can be used as a scaling parameter for characterizing the
self-similarity of the escape process.
\\ \indent An analytical expression for $T_1$ is obtained by applying
the binomial formula to the integral of Eq.~(\ref{eq:newfrompre}):
\begin{eqnarray}
T_1(N,L_c,L_0,\Gamma)
&=& 
\sum_{k=1}^N \left\{
(-1)^{k+1}
\;\binom{N}{k}\;\cdot
\right.
\label{eq:newT11}
\\&& 
\left.
\cdot\int_0^{+\infty}dt\;[S_1(t|1,L_c,L_0,\Gamma)]^k
\right\}.
\nonumber
\\&=&
\frac{L_c^2}{D_1}
\;g\left(N,L_c,L_0,\Gamma\right),
\nonumber
\end{eqnarray}
where:
\begin{eqnarray}
g(N,L_c,L_0,\Gamma)
&\equiv& 
\sum_{k=1}^N \left\{
(-1)^{k+1}
\;\binom{N}{k}\;\cdot
\right.
\label{eq:newg}
\\&& 
\left.
\cdot\int_0^{+\infty}dt'\;[S_1(t'|1,L_c,L_0,\Gamma)]^k
\right\},
\nonumber
\end{eqnarray}
and the change of variable to adimensional time $t'\equiv
  t\,D_1/L_c^2$ has been performed .
The interesting feature of Eq.~\eqref{eq:newT11} is that it expresses 
a collective behavior of $N$ particles in single-file (l.h.s.) only in 
terms of the survival probability of a single particle in the channel (r.h.s.).
Importantly, $T_1$ scales inversely with the diffusion coefficient $D_1$
of the last particle to leave the microfluidic channel. 

To validate this theoretical description, we experimentally measure
the mean escape time $T_1$ and the diffusion coefficient $D_1$ of the
last colloidal particle in the channel, after all the other particles
have left the channel. In this way, the measurement of $D_1$ (reported
in Table S1 and Fig. S1C) is not affected by particle-particle
interactions.  We evaluate $D_1$ through the mean square displacement
($MSD$) as $D_1=MSD(n \Delta t) /(2 \;n\Delta t)$ where $\Delta t$ is
the lag time between consecutive frames and $n=1,2,\ldots,N_t-1$ are
the last particle trajectory points. In agreement with previous
findings~\cite{DettP:2014, PaglE:2014}, due to particle-channel
interactions and hydrodynamic effects, $D_1$ is found to be lower  on average of about 1/5 than the diffusion coefficient measured in the bulk (Table
S1). 
With 
experimental conditions very similar to the ones used in the
present study, it has also been shown~\cite{DettP:2014} that $D_1$ 
is approximately uniform
throughout the entire channel length, the
transition region between bulk and channel being located beyond the channel ends. Furthermore, in the Supplementary Material~\cite{Loca:2015} we outline a simple theoretical discussion on the impact that particle-channel interactions have on $D_1$ at the channel entrances.

Under small external biases $F_e < 10^{-15}\;{\rm N}$, the theoretical
estimations of $T_1$ calculated according to Eq.~\eqref{eq:newT11}
(mathematical symbols in Fig. 3A) slightly overestimate the
experimental values (full symbols).  In such cases, a better
description of the experimental findings is given by an effective
theory that accounts for excluded volume effects between colloidal
particles.  In the limit of weak bias $\Gamma L_c \ll 1$ (or $k_B T
\gg F_e L_c$), the single-particle survival probability only depends
on the ratio $L_0/L_c$ and Eq.~\eqref{eq:newT11} simplifies into
\begin{equation}
T_1(N,L_c,L_0)
= 
\frac{L_c^2}{D_1}
\;h\left(N,\frac{L_0}{L_c}\right),
\label{eq:newMET1}
\end{equation}
where:
\begin{eqnarray}
h\left(N,\frac{L_c}{L_0}\right)
&\equiv& 
\sum_{k=1}^N \left\{
(-1)^{k+1}
\;\binom{N}{k}\;\cdot
\right.
\label{eq:newh}
\\&& 
\left.
\cdot\int_0^{+\infty}dt'\;\left[S_1\left(t'\left|1,\frac{L_c}{L_0}\right.\right)\right]^k
\right\}.
\nonumber
\end{eqnarray}
When $n$ particles of radius $R$ are within the channel, its length
$L_c$ reduces to: $ L_n=L_c-2\,(n-1)\,R.  $ An effective length
$L_{eff}$ can be estimated through the weighted average
\begin{eqnarray}
&L_{eff}&(N,L_c,L_0,\Gamma,R)=
\nonumber\\&=&
\frac{\sum_{n=1}^N[T_{n}-T_{n+1}]\;(L_c-2(n-1)R)}{T_1},
\end{eqnarray}
where:
\begin{equation}
T_n(N,L_c,L_0,\Gamma)=
\int_0^{+\infty}dt\;S_n(t|N,L_c,L_0,\Gamma)
\end{equation}
is the mean first passage time of the first $n\leq N$ particles
exiting the channel and $[T_{n}-T_{n+1}]$ is the average time span in which $n$ particles are found in the channel. Excluded-volume corrections  are thus effectively
taken into account by substituting $L_c$ with $L_{eff}$ in
Eq.~\eqref{eq:newMET1}.  For $F_e\,L_c\ll k_BT$ we thus obtain the
following expression:
\begin{equation}
\label{eq:newMET2}
T_1(N,L_c,L_0,R)=
\frac{L_{eff}(N,L_c,L_0,R)^2}{D_1(R,\Phi)}
\;h\left(N,\frac{L_0}{L_c}\right)
\end{equation}
The values of $T_1$ calculated according to
  Eq.~(\ref{eq:newMET2}) (open symbols in Fig. 3) are smaller than those
  obtained by using the point-like particle approximation in Eq.~\eqref{eq:newT11}
  (mathematical symbols).  Intuitively, the effective channel
  length available to each particle decreases when the particle excluded volume
  is taken into
  account.
\begin{figure}
\centering
\includegraphics[width=8.5 cm]{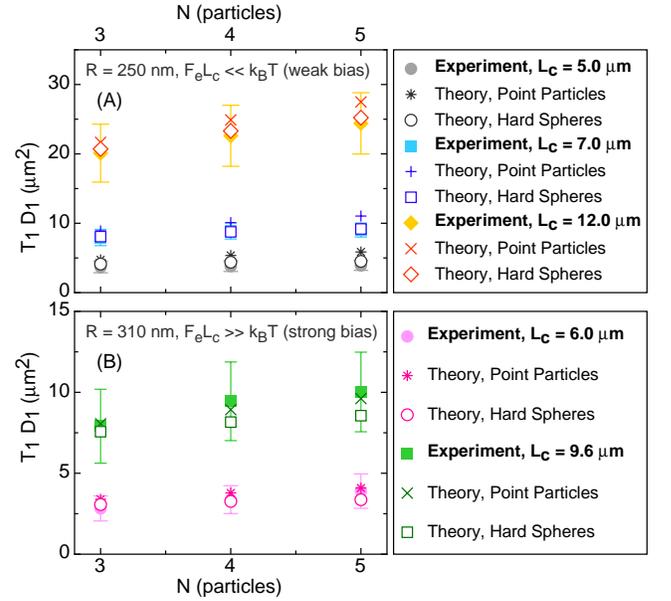}
\caption{(Color online). Dependence of the product of the mean escape
  time $T_1$ and the diffusion coefficient $D_1$ on the initial number
  $N$ of particles in the channel under weak (A) and strong (B) bias
  conditions. Full symbols are experimental data. Error bars are
  the propagation of the standard errors for $T_1$ and $D_1$ recorded
  in at least 50 experimental independent measurements. Mathematical
  and open symbols report analytical predictions calculated
  according to Eq.~(\ref{eq:newMET1}) and Eq.~(\ref{eq:newMET2}), respectively.}
\label{fig:fig3}
\end{figure}

Eq.~(\ref{eq:newMET2}) provides a more accurate description
(Fig.~\ref{fig:fig3}A, open symbols) of the experimental data (full
symbols) in the presence of a weak external bias with respect to the
point-like particle description (mathematical symbols). Consistently with the Debye
  range, we define collisions as those events for
  which the distance between the centers of two neighboring
  particles becomes smaller than $2.1 R$. Remarkably, we show that
an external force $F_e = 10^{-15}\;{\rm N}$ is sufficient to
drastically decrease the experimentally measured number of interparticle collisions (Fig. S3).   
Indeed, such a force drags all the particles in one direction reducing the collision probability. For this reason, in the presence of a 
strong bias, the point-like particle predictions are closer
to the experimental findings than those of the 
effective theory with excluded volume
effects (Fig.~\ref{fig:fig3}B).
Noteworthy, all the above
comparisons between experimental data and theoretical predictions are
carried out without fitting parameters, and are only based on the
the experimental
measurements of $L_0$ (Fig. S2), $L_c$, $D_1$, and 
$F_e$ (see Table S1).  

\indent In summary, we have found that the escape process of
  $N$ particles in single-file can be described in terms of the
  survival probability of the last particle to leave the channel and that
  the mean escape time of the process scales inversely
  with the diffusion coefficient of such particle. By demonstrating that it is sufficient to investigate the diffusion of a single
  particle, our findings streamline the design of synthetic
  arrays of channels and pores for applications such as filtering where 
  multi-particle transport under close confinement is paramount. By proving that the simple formula $F_e\,L_c\simeq k_BT$,
  ($F_e\simeq 1\;{\rm fN}$ in our experiments) can be
  exploited to quantify the effect of an external force on the escape process, our findings help both rationalizing metabolites and drugs diffusion across biological membranes under an external force and optimizing device geometry in applications involving particle transport in a force gradient. 
 In this respect our theoretical model offers a novel framework for (\textit{i}) investigating the escape
  process from widespread biological and synthetic 
  constrictions, and (\textit{ii}) providing guidelines for the design of
  micro- and nano-devices for particle and molecule transport applications.  \\
\begin{acknowledgments}
The authors are indebted to Dr U.F. Keyser for helpful discussions. S.P. acknowledges the support from the Leverhulme Trust through an Early Career Fellowship (ECF-2013-444), the Wellcome Trust through an Institutional Strategic Support Fund (SW-05377) and the Royal Society through a Research Grant (RG140203). M.P. and E.L. kindly acknowledge funding from the European Research Council under the European Community's Seventh Framework Programme (FP7/2007-2013)/ERC Grant Agreement No. 279004.
\end{acknowledgments}

\bibliography{Biblio}

\end{document}